\documentclass[fleqn]{jaes}

\jyear{2023}
\jmonth{Nov.}

\usepackage{amsmath}
\usepackage{amsmath,graphicx,amssymb,bm,bbm}
\usepackage{bm}
\usepackage{algorithm}
\usepackage{algpseudocode}
\usepackage{hyperref}
\usepackage{booktabs}
\usepackage{lipsum}
\usepackage{svg}
\usepackage{draftwatermark}
\SetWatermarkText{}
\SetWatermarkColor[gray]{0.9}
\SetWatermarkScale{1.4}

\begin{document}

\markboth{MOLINER AND VÄLIMÄKI}{DIFFUSION-BASED AUDIO INPAINTING}

\title{Diffusion-Based Audio Inpainting}

\authorgroup{
\author{ELOI MOLINER}
AND \author{VESA VÄLIMÄKI,}
\role{AES Fellow}
\email{(eloi.moliner@aalto.fi)\quad\quad\quad\quad\quad\quad\quad\quad\quad\quad\quad (vesa.valimaki@aalto.fi)}
\affil{Acoustics Lab, Department of Information and Communications Engineering, Aalto University, Espoo, Finland}
}

\abstract{%
Audio inpainting aims to reconstruct missing segments in corrupted recordings. Most existing methods produce plausible reconstructions when the gap lengths are short, but struggle to reconstruct gaps larger than about 100\;ms. This paper explores diffusion models, a recent class of deep learning models, for the task of audio inpainting. The proposed method uses an unconditionally trained generative model, which can be conditioned in a zero-shot fashion for audio inpainting, and is able to regenerate gaps of any size. An improved deep neural network architecture based on the constant-Q transform that allows the model to exploit pitch-equivariant symmetries in audio is also presented. The performance of the proposed algorithm is evaluated through objective and subjective metrics for the task of reconstructing short to mid-sized gaps, up to 300\,ms. The results of a formal listening test indicate that, for short gaps in the range of 50\,ms, the proposed method delivers performance comparable to the baselines. For wider gaps up to 300\,ms long, our method outperforms the baselines and retains good or fair audio quality.  The method presented in this paper can be applied to restoring sound recordings that suffer from severe local disturbances or dropouts.
}

\maketitle


\section{INTRODUCTION}

Audio inpainting refers to repairing or filling in missing or degraded parts of an audio signal \cite{adler2011audio}. Inpainting can be used to remove noise, glitches, or other unwanted artifacts from an audio recording, or to fill in missing sections of audio that have been lost or damaged.
Application examples include the restoration of old recordings corrupted by local disturbances \cite{godsill2013digital, ragano2022automatic}, the reconstruction of missing audio samples caused by scratches in CDs \cite{janssen1986adaptive}, or compensation for audio packet loss in communication networks \cite{goodman1986waveform}. In addition, audio inpainting can be used in music and audio production to create special effects or to manipulate audio signals in creative ways \cite{bazin2021spectrogram}. This paper presents a novel audio inpainting method based on  diffusion models \cite{ho2020denoising,song2020score}, a recently proposed generative deep-learning technique.

The task of audio inpainting is an ill-posed inverse problem, characterized by a non-unique set of solutions.
 Audio inpainting has been widely studied in the literature \cite{adler2011audio,lieb2018audio, marafioti2019context, mokry2020audio, taubock2020dictionary}. 
The methods employed in audio inpainting are primarily distinguished by the way the observed signal samples are used as a prior, or how they incorporate pre-existing assumptions about the signal.
For instance, some techniques are based on autoregression \cite{janssen1986adaptive} or signal sparsity \cite{mokry2019introducing}.
However, most of these techniques demonstrate strong performance only when applied to gaps of less than 100\,ms in duration. Such techniques tend to encounter challenges with longer gaps or in cases where the assumption of stationarity, explicitly required by autoregressive methods and implicitly relied upon by sparsity-based methods, does not hold.

In this work, we use generative priors, learned from a diffusion probabilistic model, assuming that the solution belongs to the same probability distribution as the dataset used for training.
Inpainting methods based on deep generative models can reach new levels of expressivity, since they are not grounded by the stationarity condition and can generate new events in the inpainted gap \cite{ebner2020audio,marafioti2020gacela,
moliner2022solving}. 
In particular, diffusion models have a strong potential to excel at this task as they possess a great versatility for solving inverse problems \cite{kawar2022denoising,chung2022diffusion, moliner2022solving}.

 In our previous study \cite{moliner2022solving}, the invertible Constant-Q Transform (CQT) was used with a diffusion model to solve inverse problems in audio.
This paper revisits the use of the CQT, proposing an improved neural network architecture operating in the transform domain using a small amount of signal redundancy. A diffusion model, built with a deep neural network, is first trained with audio material as an unconditional generator. During inference, the model is conditioned in a zero-shot manner to generate a plausible reconstruction of the missing segment. In contrast to existing audio inpainting methods \cite{marafioti2019context, mokry2020audio, taubock2020dictionary}, the proposed diffusion model can restore gaps of arbitrary length, retaining high quality for longer gaps.

This paper addresses the inpainting of compact gaps in an audio signal without any accompanying side information. Specifically, we focus on gaps in the short-to-medium size range, ranging from 25 to 300\,ms. 
Note that this differs from 
the goal of our previous work \cite{moliner2022solving}, where the model was tested on larger gaps up to 1.5\,s in length.
We observed that when the gap was very long, the model had to generate new events. Although these generated events were often statistically plausible, they did not align with the musical context and were deemed musically incorrect, which is undesirable. This led us to conclude that a practical inpainting method for large gaps would require a high-level understanding of the music structure or the ability to be conditioned with a guiding signal, as proposed in recent research \cite{liu2023maid}. However, such considerations fall outside the scope of this paper.
As a result, we limit the evaluation to gaps no longer than 300\,ms. Within this range, we assume that the content to be filled can 
be anticipated by a human listener, ensuring a reliable evaluation of the inpainting performance.

The rest of this paper is organized as follows. Sec.~\ref{overview} reviews the relevant audio and image inpainting literature. Sec.~\ref{diffusion} explains the basics of diffusion models and the conditioning method for the inpainting task. Sec.~\ref{architecture} introduces the new diffusion-model architecture, which employs the invertible CQT. Sec.~\ref{evaluation} compares the proposed method with previous inpainting methods in terms of objective and subjective metrics. Sec.~\ref{conclusion} concludes the paper.

\section{OVERVIEW OF INPAINTING METHODS} \label{overview}

 This section reviews some relevant methods in the audio inpainting literature.
 In addition, we summarize some recent work on image inpainting with diffusion models that inspired this work.
 
\subsection{Existing Audio Inpainting Methods} 
 Adler et al. first used the term “audio inpainting” to describe the restoration of gaps in audio signals \cite{adler2011audio}, adopting the name from the image inpainting literature. However, this is an old problem in audio processing, and the same task has been previously referred to in the literature as audio interpolation \cite{janssen1986adaptive, etter1996restoration,esquef2003interpolation}, audio extrapolation \cite{kauppinen2001method, kauppinen2002audio}, reconstruction of missing samples \cite{kauppinen2002reconstruction, esquef2006}, waveform substitution \cite{goodman1986waveform}, and imputation \cite{smaragdis2009missing}, among other things. 
The first methods used interpolation techniques based on the observed samples surrounding the gap \cite{janssen1986adaptive}. A family of  successful methods uses autoregressive modeling based on the assumption that the signal is stationary and can be approximated by a linear combination of past samples \cite{janssen1986adaptive,etter1996restoration,esquef2003interpolation}.

 A more recent family of methods takes advantage of sparse signal representations \cite{lieb2018audio, mokry2020audio, mokry2019introducing}, such as the short-time Fourier transform (STFT).
 These methods try to find the sparse representation of the missing 
 part of the signal that best fits the surrounding, uncorrupted signal.
An established method to enhance sparsity-based audio inpainting is to learn the dictionary of basis functions \cite{taubock2020dictionary, rajbamshi2021audio}. Another recent work uses non-negative matrix factorization, exploiting the low-rankness of the magnitude spectra as a prior \cite{mokry2022algorithms}.

The methods mentioned above only perform well for inpainting short gaps, roughly in the range from 10\,ms to 100\,ms. For longer gaps, these methods tend to fail to produce plausible reconstructions since the stationarity condition does not hold true.
Some inpainting attempts for long gaps are based on strong assumptions about the underlying structure of the gap, including sinusoidal modeling
\cite{lagrange2005long} or similarity graphs
\cite{perraudin2018inpainting}.

\subsection{Deep-Learning Based Audio Inpainting}

During the last few years, a new trend has emerged using deep-learning-based techniques for audio restoration, including the task of inpainting.
Most of these studies use generative models as the prior for inpainting. This allows for methods that are able to generate new content in the gaps to be filled. 
For instance, Generative Adversarial Networks (GANs) have been explored for this task \cite{ebner2020audio,  marafioti2020gacela, greshler2021catch}. 
Most of these methods are based on a supervised problem-specialized setting, where a dataset of masked/reconstructed audio signals needs to be built to train the model. 
A shortcoming of this approach is that a model trained  with a certain set of degradations does often not generalize to unseen degradations and, as a consequence, lacks the versatility to be applied for restoring gaps of arbitrary length.

Some other closely related studies fall under the category of packet-loss concealment, which is a similar problem to audio inpainting but with real-time constraints and usually targeting speech signals. Within this context, predictive methods based on convolutional and recurrent neural networks \cite{lee2015packet, lin2021time} as well as GANs \cite{ pascual2021adversarial,ou2022concealing} have been proposed.

Also worth mentioning are other recent works that have applied multi-modal side information as a conditioner for the inpainting algorithm, including video frames \cite{morrone2021audio}, symbolic music \cite{cheuk2022diffroll, liu2023maid}, or text \cite{borsos2022speechpainter, wang2023audit}. Although this idea falls outside the scope of this paper, exploiting multi-modal information may turn out to be beneficial to inpaint large gaps, where the context of the gap does not contain enough information to reconstruct the missing segment.

\subsection{Diffusion Models for Image Inpainting}

Deep generative models have tremendously impacted image processing research, not least their application to the image inpainting problem.
Relevant to this paper are recent papers applying diffusion models for image inpainting.
There are two main strategies in the literature to solve inverse problems with diffusion models, including inpainting. 
The first one consists of sequentially replacing the observed part of the signal in the reverse diffusion process  \cite{song2020score, wang2022zero}. 
This idea ensures data consistency and is conceptually simple but, in practice, struggles to generate consistent content.
Other work refined this approach by incorporating ideas that benefited its versatility and performance, such as using singular-value decomposition \cite{kawar2022denoising} or multiple resampling during each sampling step \cite{lugmayr2022repaint}.
The other strategy builds on a Bayesian interpretation of posterior sampling and estimates the gradients of the log-likelihood function \cite{chung2022diffusion, song2022pseudoinverse}, as elaborated in Sec.~\ref{posterior_sampling}. 
These methods allow for a good approximation of the posterior distribution, which leads to enhanced inpainting results, at the expense of a higher computational cost.

\section{A DIFFUSION MODEL FOR AUDIO INPAINTING} \label{diffusion}

Diffusion models are a class of generative models that have gained interest during recent years for a wide range of modalities, such as images \cite{ho2020denoising, dhariwal2021diffusion, karras2020analyzing}, audio \cite{kong2020diffwave, richter2022speech,moliner2022solving}, video \cite{ho2022video}, and symbolic music \cite{cheuk2022diffroll}, among others.
These models generate new data instances by reversing the diffusion process, by which data $\bm{x}_0 \sim p_\text{data}$ is progressively diffused into Gaussian noise $\bm{x}_{T} \sim \mathcal{N}(\mathbf{0},\sigma_\text{max}^2 \mathbf{I})$ over time $\tau$ \cite{ho2020denoising}\footnote{The ``diffusion time'' variable $\tau$  must not be confused with the ``audio time'' $t$. We use this formulation for notational consistency.}.

We follow the parameterization by Karras et al. \cite{karras2022elucidating}, who define the reverse diffusion process with the following \textit{probability flow ordinary differential equation (ODE)}:
\begin{equation}\label{odekarras}
    \text{d}\bm{x}=  - \dot\sigma(\tau) \sigma(\tau)  \nabla_{\bm{x}}\log p_\tau(\bm{x}_\tau) \text{d}\tau,
\end{equation}
where $\text{d}\tau$ is an infinitesimal negative timestep, the noise level is defined as $\sigma(\tau)=\tau$, and its first derivative as $\dot\sigma(\tau)=1$.
The ODE is governed by the gradient of the log probability density $\nabla_{\bm{x}}\log p_\tau(\bm{x}_\tau)$, formally known as the \textit{score} function  \cite{hyvarinen2005estimation}.

The score is analytically intractable but can be approximated as
\begin{equation}
    \nabla_{\bm{x}}\log p_\tau(\bm{x}_\tau) \approx (D_\theta(\bm{x}_\tau,\tau)-\bm{x}_\tau)/\sigma(\tau)^2,
\end{equation}
where $D_\theta(\bm{x}_\tau, \tau) = \bm{\hat{x}}_0$, which refers to $\hat{\mathbf{x}}_\tau$ at $\tau=0$, is a deep neural network with weights $\theta$, optimized with a denoising Euclidean objective:
 \begin{equation}\label{loss}
    \mathbb{E}_{\bm{x}_0 , \boldsymbol\epsilon \sim \mathcal{N}(\mathbf{0},\mathbf{I}) }  \left[ \lambda(\tau) \lVert D_\theta(\bm{x}_0+\sigma(\tau)\bm{\epsilon},\tau) -\bm{x}_0   \rVert_2^2 \right],
\end{equation}
where $\lambda(\tau)$ is a time-dependent weighting parameter. Furthermore,
we use the preconditioning strategy proposed by Karras et al. \cite{karras2022elucidating}:
\begin{equation} \label{denoise}
D_\theta(\bm{x}_\tau, \tau)=
c_\text{skip}(\tau)\bm{x}_\tau+
c_\text{out}(\tau)F_\theta(c_\text{in}(\tau)\bm{x}_\tau, \tau),
\end{equation}
where $F_\theta$ represents an optimizable deep neural network, and $c_\text{skip}(\tau)$, $c_\text{out}(\tau)$, and $c_\text{in}(\tau)$ are weighting parameters optimized in such a way that the input and the output of $F_\theta$ always have close-to-unit variance, a well-known good practice when training deep neural networks.

For more comprehensive details on the diffusion model formalism and optimization, as well as the optimal weighting parameters, we refer to \cite{karras2022elucidating}.
 In the rest of this section, we elaborate on the audio inpainting problem and the required changes that are applied to the inference process of a diffusion model to solve this task.

\subsection{Inverse Problem Formulation}
The audio inpainting task can be formulated as a linear inverse problem \cite{adler2011audio}. Consider an audio signal $\bm{x}_0$ and its observed version $\bm{y}$ with missing samples. We can write their relation as
\begin{equation}
    \bm{y}=\textbf{m}\odot\bm{x}_0,
\end{equation}
\noindent where $\textbf{m}$ is a binary mask operator and $\odot$ represents the Hadamard product, or element-wise multiplication. 
 In this work, we consider the operator $\textbf{m}$ as a known compact binary mask, having the value 0 at locations where samples are missing and 1 otherwise. The goal is to recover the original signal $\bm{x}_0$ when the observed measurements $\bm{y}$ and mask $\textbf{m}$ are known.

\subsection{Audio Inpainting via Posterior Sampling} \label{posterior_sampling}

The iterative nature of diffusion models offers great flexibility for solving inverse problems \cite{chung2022diffusion, moliner2022solving}. All that is needed is to substitute the score in Eq.~\eqref{odekarras} with the \textit{posterior score} $\nabla_{\bm{x}}\log p_\tau(\bm{x}_\tau|\bm{y})$ \cite{song2020score}. 

Applying Bayes' rule, the posterior factorizes as $p_\tau(\bm{x}_\tau|\bm{y}) \propto p_\tau(\bm{x}_\tau) p_\tau(\bm{y}|\bm{x}_\tau)$, which leads to
 \begin{equation} \label{posterior}
\nabla_{\bm{x}_\tau}\log p_\tau(\bm{x}_\tau|\bm{y})=
 \nabla_{\bm{x}_\tau}\log p_\tau(\bm{x}_\tau)+
     \nabla_{\bm{x}_\tau}\log p_\tau(\bm{y}|\bm{x}_\tau),
\end{equation}
We refer to the term $\nabla_{\bm{x}_\tau}\log p_\tau(\bm{y}|\bm{x}_\tau)$ as the \emph{noise-perturbed likelihood score}. Note that this term cannot be derived in closed form due to its dependence on the noise level $\sigma(\tau)$, since $\bm{x}_\tau$ represents the noise-perturbed signal $\bm{x}_\tau=\bm{x}_0 + \sigma(\tau) \bm{\epsilon}$, where $\bm{\epsilon}\sim \mathcal{N}(\mathbf{0},\mathbf{I})$. However, Chung et al. \cite{chung2022diffusion} propose to approximate the \textit{noise-perturbed likelihood} with $p_\tau(\bm{y}|\bm{x}_\tau) \simeq p(\bm{y}|\hat{\bm{x}}_0)$,
where $\hat{\bm{x}}_0$ is the denoised estimate at an intermediate noise level.

Modeling the likelihood as a normal distribution,  the \textit{noise-perturbed likelihood score} is approximated as
\begin{equation}\label{recguid}
    \nabla_{\bm{x}}\log p_\tau(\bm{y}|\bm{x}_\tau) \simeq
    -\xi(\tau) \; \nabla_{\bm{x}} \lVert \bm{y} - \textbf{m}\odot\bm{\hat{x}}_0
 \rVert^2.
\end{equation}
This strategy can be understood as a sort of guidance \cite{ho2022video}, in analogy with classifier guidance \cite{dhariwal2021diffusion}. 
Importantly, note that the gradient computation requires differentiating through the neural network $F_\theta$, which is responsible for the estimation of $\bm{\hat{x}}_0$, resulting in computational overhead.

The variable $\xi(\tau)$ in Eq.~\eqref{recguid} is a scaling function that defines the amount of guidance that is applied during sampling or, in other words, how strongly the conditioning affects the sampling trajectories.
We parameterize the scaling function as \cite{moliner2022solving}
 \begin{equation}
\xi(\tau)=\xi^\prime \sqrt{N}/( \sigma(\tau) \lVert \nabla_{\bm{x}} \lVert \bm{y} - \textbf{m}\odot\bm{\hat{x}}_0 \rVert^2 \rVert^2),
 \end{equation}
 where $N$ is the length of the audio signal in samples and $\xi^\prime$ is a scalar hyperparameter. Choosing $\xi^\prime=0$ leads to an unconditional sampler, but selecting too large a value for $\xi^\prime$ results in a degenerate solution. 
 This parameterization scales the likelihood gradient by its norm in a way similar to \cite{kim2021guided}, regularizing the influence of the likelihood throughout the inference process.
 We empirically observed through qualitative analysis that this strategy allows for robust results.


However, the above conditioning method does not ensure data consistency with the observed samples. When the observed samples $\bm{y}$ are noiseless and reliable, as we assume in this work, the preferred solution is to keep them unchanged in the final output. 
A straightforward way to avoid changing the existing samples is replacing the reliable samples from the intermediate estimates $\bm{\hat{x}}_0$ using the inpainting mask.
To keep the observed samples, the following data consistency step at each sampling iteration can be applied:
\begin{equation}\label{consistencystep}
    \bm{\hat{x}}_0^\prime = \bm{y} +(\bm{1}-\textbf{m})\odot\bm{\hat{x}}_0.
\end{equation}

Although some studies have proved the data consistency step suboptimal \cite{chung2022diffusion, chung2022improving}, others rely solely on data consistency as a method to condition the diffusion model to solve inverse problems \cite{song2020score, lugmayr2022repaint}.
We observe that applying data consistency steps usually produces discontinuity effects at the boundaries of the mask.
 To mitigate this effect, we apply a smoothed version of the mask $\textbf{m}$ for the data consistency step of Eq.~\eqref{consistencystep}, which is implemented by fading 1\,ms of the reliable signal at the edges of each gap with a raised cosine function. 

\subsection{Inference}
\begin{figure*}[t]
    \centering
    \includegraphics[width=0.94\textwidth, trim={0 0.04cm 0 0.1cm}, clip]{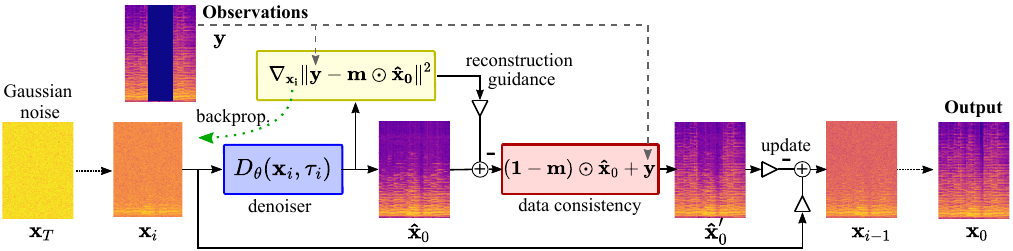}
    \caption{
    Inference block diagram for audio inpainting, where all straight lines represent a feedforward signal flow in the time domain. The deep neural network is included in the denoiser block. 
    The computation of the reconstruction gradient requires differentiating through the mask and the denoiser block by means of backpropagation, denoted as ``backprop." above, requiring a backward pass through the deep neural network, illustrated here with a  dotted line. 
    The spectrograms are shown for illustrative purposes. 
    }
    \label{fig:inference}
\end{figure*}
Having defined the probability flow ODE, Eq.~\eqref{odekarras}, and the posterior sampling mechanism, Eq.~\eqref{recguid}, the
next tasks are to discretize and solve the reverse diffusion process, using a trained diffusion model. 
In this work, we use the second-order stochastic sampler proposed by Karras et al.~\cite{karras2022elucidating}, offering a good tradeoff between algorithmic complexity and accuracy. This sampler also adds controllable stochasticity into the process, which is intended to regularize approximation errors. The sampling algorithm, specific for inpainting, is described in Algorithm 1.
Fig.~\ref{fig:inference} summarizes graphically the sampling process, omitting the second-order correction for brevity.
It is important to note that implementing the second-order correction would necessitate an additional denoising forward pass and the guidance computation.

As presented on the left-hand side of Fig.~\ref{fig:inference}, the audio signal $\bm{x}_T$ is initialized with Gaussian noise, and the noise is iteratively removed throughout the inference process. 
During each discretization step, we acquire a denoised estimate denoted as $\bm{\hat{x}}_0$. We then condition this estimate with the input observations shown at the top of Fig.~\ref{fig:inference} by incorporating the reconstruction gradient from Eq. \eqref{recguid} and implementing the data consistency step described in Eq. \eqref{consistencystep}.
Each update step is a weighted sum of the noisy signal at the given step $\bm{x}_i$ and the modified denoised estimate $\bm{\hat{x}}_0^\prime$. At the end of the process, shown on the right-hand side of Fig.~\ref{fig:inference}, the noise level becomes imperceptible and, thus, the reconstructed output signal $\hat{\bm{x}}_0$ is obtained.

\begin{algorithm}[t]
\caption{Inference Conditioned for Audio Inpainting}
\label{alg3}
\begin{algorithmic}
\Require observations $\bm{y}$, inpainting mask $\bm{m}$, number of iterations $T$, noise schedule $\tau_i$, stochasticity  $S_\text{churn}$
\State Sample $\bm{x}_{T}\sim \mathcal{N}(\mathbf{0},\sigma_\text{max}^2\mathbf{I})$ \Comment{Initial noise is $\bm{x}_T$}
\State $\gamma = \min(S_\text{churn}/T, \sqrt{2} -1) $ \Comment{Amount of stochasticity}
\For{$i=T, \dots, 1$} \Comment{Step backwards}
\State $\tilde{\tau}_i=(1+\gamma)\tau_i $ \Comment{Increased noise level}
\State Sample $\bm{\epsilon} \sim \mathcal{N}(\mathbf{0}, \mathbf{I})$ 
\State $\tilde{\bm{x}_i}=\bm{x}_i+\sqrt{\sigma(\tilde{\tau}_{i})^2 - \sigma(\tau_{i})^2} \bm{\epsilon}$ \Comment{Add extra noise}
\State $\hat{\bm{x}}_0= D_\theta(\tilde{\bm{x}_{i}}, \tilde{\tau}_{i})$ \Comment{Denoiser}
\State $\hat{\bm{x}}_0=H_\text{post}(\hat{\bm{x}}_0)$ \Comment{Post-processing filter}
\State $\hat{\bm{x}}_0= \hat{\bm{x}}_0 - \sigma(\tilde{\tau}_i)^2 \xi(\tilde{\tau_i})  \; \nabla_{\tilde{\bm{x}}} \lVert \bm{y} -
\textbf{m}\odot\hat{\bm{x}}_0 \rVert^2 $ \Comment{Eq.~\eqref{recguid}}
\State $\hat{\bm{x}}_0^\prime=\bm{y}+ (\bm{1}-\textbf{m})\odot\hat{\bm{x}}_0$\Comment{Data consistency}
\State $\bm{x}_{{i-1}} = \tilde{\bm{x}_{i}} - (\sigma(\tau_{i-1})-\sigma(\tilde{\tau}_i)) \left(\frac{\hat{\bm{x}}_0^\prime- \tilde{\bm{x}_i}}{\sigma(\tilde{\tau}_i)}\right)$ \Comment{Update step}
\EndFor
\State \Return $\hat{\mathbf{x}}_0$  \Comment{Output is the reconstructed signal}
\end{algorithmic}
\end{algorithm}

The noise schedule represents one of the most critical design choices.
Also following \cite{karras2022elucidating}, given a number of discretization steps $T$, we define the noise levels as
\begin{equation}\label{schedule}
    \tau_{i}=\left(\sigma_{\text{max}}^{\;\;\;\frac{1}{\rho}} 
    + \tfrac{i}{T-1}\left(
    \sigma_\text{min}^{\;\;\;\frac{1}{\rho}}
    -\sigma_\text{max}^{\;\;\;\frac{1}{\rho}}
    \right)\right)^\rho,
\end{equation}
where $0\leq i\leq T-1$ is the discretization index, $\sigma_\text{min}$ and $\sigma_\text{max}$ are, respectively, the minimum and maximum noise levels, and $\rho$ is a parameter controlling the warping of the schedule, with higher values of $\rho$ representing more steps at lower noise levels. As in \cite{moliner2022solving}, we choose $\sigma_\text{min}=10^{-4}$, $\sigma_\text{max}=1$, and $\rho=13$. The number of steps $T$ exhibits a clear tradeoff between sample quality and speed. We use the value $T=70$ in our experiments.

The amount of stochasticity injected into the process
 is controlled with the parameter $S_\text{churn}$ \cite{karras2022elucidating}.
Empirically, we have observed that adding a certain amount of stochasticity helps to produce clean outputs. In our experiments, we choose $S_\text{churn}=10$.

\section{IMPROVED CQT-BASED ARCHITECTURE} \label{architecture}

Diffusion models are architecture agnostic, imposing no constraints on how the denoiser backbone architecture $F_\theta$ is designed.
However, although the choice of architecture does not have theoretical implications, in the best case 
it can accelerate the convergence of the diffusion, producing perceptually-satisfying samples efficiently.

In our previous study \cite{moliner2022solving},
we used an invertible CQT \cite{velasco2011constructing} to leverage structure from the audio signal and to exploit the pitch-equivariant symmetry that harmonic signals exhibit when they are represented in a logarithmically-spaced time-frequency domain. 
The most interesting property of the CQT is that a translation on the frequency axis is equivalent to pitch transposition. This symmetry motivates the usage of two-dimensional convolutional neural networks (CNNs), considering that the convolutional operator, which CNNs are composed of, is translation equivariant.
In this section, we elaborate the usage of a CQT and introduce an improved neural network architecture that 
processes audio signals as CQT spectrograms. We call the improved diffusion model based on this architecture the CQT-Diff+ algorithm. 

\subsection{Using a CQT Representation}

The diffusion process described in Sec.~\ref{diffusion} is developed in the time domain. However, as part of the computation inside the deep neural network $F_\theta$, the input waveform is represented with an invertible CQT. 
Concisely, the neural network $F_\theta$ is composed as
\begin{equation}
F_\theta =\text{ICQT} \circ F_\theta^\prime \circ \text{CQT},
\end{equation}
where CQT and ICQT are the constant-Q-transform operation and its inverse respectively, $\circ$ is the function composition operation, 
and $F_\theta^\prime$ refers to the neural network layers with trainable weights.
This approach takes advantage of the structure imposed by the CQT while maintaining maximum versatility.
Applying the neural network weights in the transform domain does not impact the optimization, since both the forward transform and its inverse are differentiable.

We use the CQT proposed by Velasco et al.~\cite{velasco2011constructing} and by Holighaus et al.~\cite{holighaus2012framework}. Briefly, this transform is built on a set of $K$ bandpass filters $g_k$ with an equal Q-factor and logarithmically-spaced center frequencies, defined as
\begin{equation}
    f_k=f_\text{min} 2^\frac{k-1}{B}, \,\,\,\text{for } k=1, 2, 3..., K,
\end{equation}
where $B$ is the number of bins per octave band (when the number of octave bands is $N_\text{oct}=K/B$) and $f_\text{min} = f_1$ is the lowest center frequency. The maximum center frequency can be designed to be placed at the Nyquist limit $f_K=f_\text{s}/2$. 
The CQT is applied using the FFT-based processing, as introduced in \cite{velasco2011constructing}, which allows for a computationally efficient implementation.
We refer to \cite{velasco2011constructing,holighaus2012framework} for further details on the CQT transform, as well as to our publicly available implementation\footnote{\href{https://github.com/eloimoliner/CQT_pytorch}{https://github.com/eloimoliner/CQT\_pytorch}}.

\subsubsection{Discarding the DC Component}
 An obvious inconvenience caused by the logarithmic frequency resolution is that there is no DC bin at 0\;Hz. If perfect reconstruction is required, one solution is to encode the DC component with a low-pass filter $g_\text{DC}$. In our prior work \cite{moliner2022solving}, the DC component was included in the model input by concatenating it to the time-frequency matrix. 
However, we observed the presence of low-frequency artifacts in the generated outputs, which we attribute to the disruption of the logarithmically uniform frequency resolution at the DC component.
Therefore, we made the decision to discard the DC component.
 By excluding the DC component, the completeness of the CQT as a transform is compromised.
 The model would now train on only a subset of the transformed space, producing an irreducible error in the frequency bands that it is blind to. However, this issue does not represent a problem in practice, since the audio signals are assumed to be bandlimited, and very little relevant information exists below 43\,Hz, which corresponds to the lowest frequency band in our specific case.
 
 
 Nevertheless, this irreducible error must be accounted for when propagating the loss during training as well as during the sampling stage. This compensation can be implemented by applying a post-processing filter to the denoiser output:
\begin{equation}
    \bm{\hat{x}}_0=H_\text{post} (D_\theta(\bm{x}_\sigma, \sigma)),
\end{equation}
where $H_\text{post}$ is a DC notch filter, designed to suppress the frequency range that is not covered by the CQT bandpass filters $g_k$. This filter is applied during both training and inference after each forward evaluation of $D_\theta(\cdot)$.


\begin{figure*}[t]
    \centering
    \includegraphics[width=0.80\textwidth]{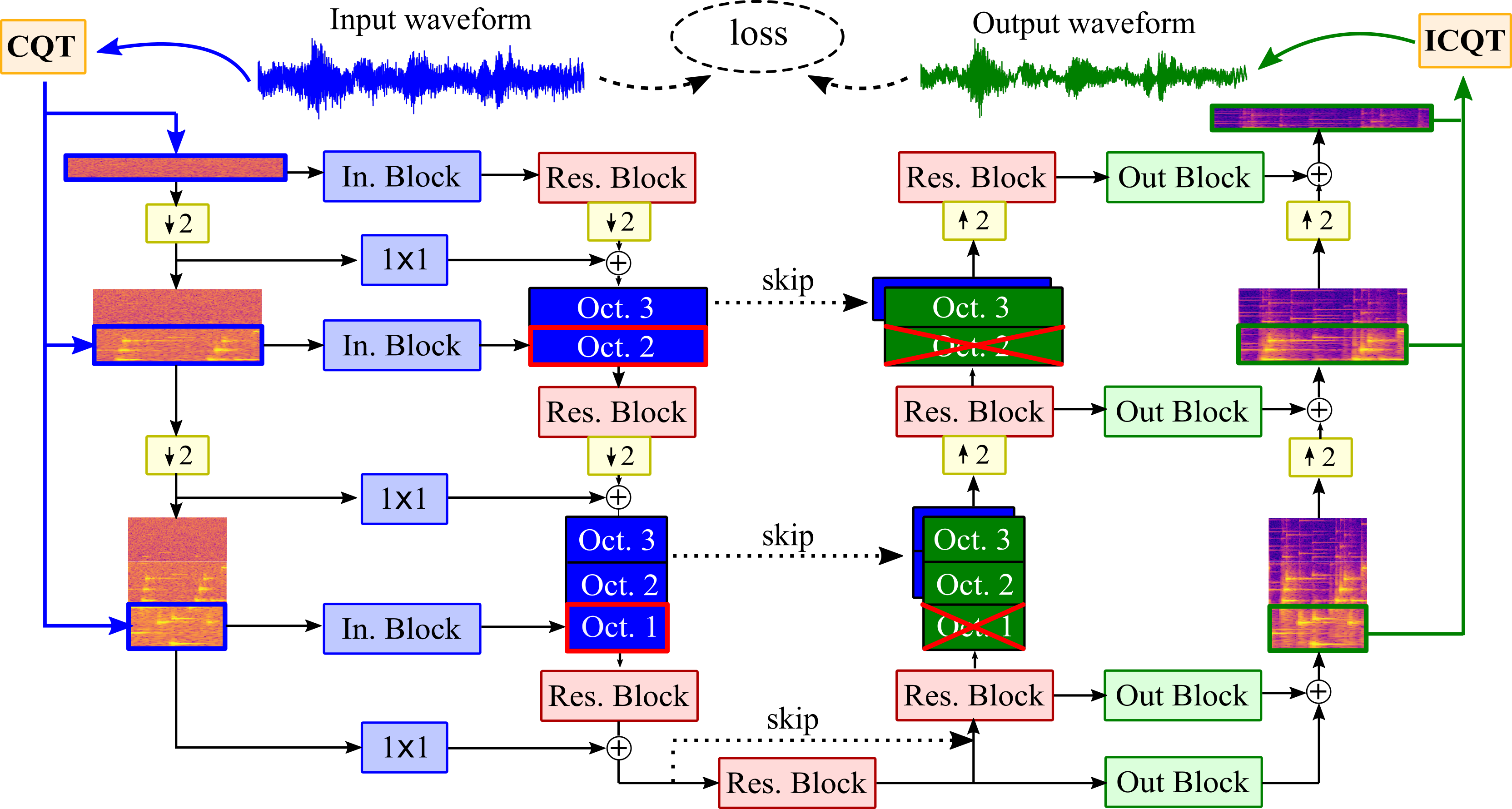}
    \vspace{-0.2cm}
    \caption{Main diagram of the CQT-U-Net deep neural network architecture. 
    In the diagram, only three octaves of eight are shown for clarity. 
    The sizes of the spectrograms are not proportional to the real signals.}
    \label{fig:architecture}
\end{figure*}

\subsubsection{Optimizing Redundancy}

In CQTs, the receptive field of the filters decreases geometrically with frequency. To guarantee invertibility, the decimation factors of the frequency bands also need to decrease geometrically, producing a non-uniform sampling grid. This feature is not only impractical for constructing a parallelizable and GPU-efficient implementation, but also complicates the architecture of the neural network.

An easy way to overcome this problem is to use instead a CQT with a uniform sampling grid, where the decimation factors remain constant across the frequency range \cite{schorkhuber2010constant}. Previously, this approach allowed treating the CQT as a 2-D matrix, and thus, to directly apply 2-D CNNs \cite{moliner2022solving}. 
A major drawback of this strategy is its overcompleteness,
which leads to a suboptimal consumption of memory and computational requirements, as the 2-D CNN is forced to process a substantial amount of signal redundancy. This considerably slowed down the training and inference processes in our previous study \cite{moliner2022solving}, and limited the potential scalability of the model.

Sch\"orkhuber et al.~\cite{schorkhuber2012pitch} proposed splitting the CQT into a sequence of octave-wise sub-transforms as a way to reduce redundancy.
We adopt this strategy, and apply different sub-transforms with constant decimation factor for each octave band, in our case each of them having 64 frequency bins.
A strong advantage of separating the CQT into octave bands is that, when powers of two are used as the sequence length, the time resolution decreases exactly by a factor of two between two consecutive octaves. This choice leads to a hierarchical representation that is suitable for processing with a U-Net architecture \cite{ronneberger2015u}. 

\subsection{Architecture Design}

A U-Net has a hierarchical encoder/decoder structure, as shown in Fig.~\ref{fig:architecture}, where the left-hand side is the encoder and the right-hand side is the decoder. The center part of Fig.~\ref{fig:architecture} is called the ``bottleneck'', which corresponds to the lowest point of the letter ``U''.
The temporal resolution is progressively reduced by a factor of two between consecutive layers in Fig.~\ref{fig:architecture}, while the frequency resolution remains unchanged \cite{moliner2022solving}.  We make use of this hierarchy by concatenating features from each CQT octave at the U-Net layers where the time resolutions match, as Fig.~\ref{fig:architecture} also illustrates.

The proposed architecture utilizes a double real representation of the complex CQT features, where the real and imaginary parts are stacked as two separate channels. Thus, the real and imaginary parts are freely merged in the channel dimension of the network, where the number of channels is further increased, but the synchrony between real and imaginary parts in the time-frequency space is conserved. 
The chosen strategy aims to circumvent the computational complexity associated with complex-valued layers, since they generally lack empirical performance advantages compared to their real-valued counterparts while imposing higher computational costs \cite{wu2023rethinking}.
However, even though the neural network views the features as real, the underlying data is complex,
and one must be cautious with how the features are processed.

In particular, we observed that shift-based operations, such as biases in convolutional layers or mean normalizations, introduced perceptual artifacts to the generated output and, as a consequence, they should be avoided.
The intuition behind this lies in the unique nature of complex numbers and the way they interact during computations.
In a complex number, the real and imaginary parts represent different dimensions of information. If shift-based operations were applied, they could introduce imbalances between the real and imaginary components, leading to inconsistent phase relationships and distorted information.
Thus, bias terms in all the layers are set to zero.
However, this does not apply to additive residual connections, since they are designed to add the activations of one layer to another without altering their phase relationships.
Note that this does not represent a practical limitation to the model because all the signals are approximately zero mean.

In accordance with typical U-Net architectures, concatenative skip connections bridge the intermediate resolutions of the encoder and decoder as shown in Fig.~\ref{fig:architecture}. An antialiasing filter was used in the downsampling and upsampling layers in the encoder and decoder stages, respectively. At each resolution of both the encoder and the decoder stages and at the bottleneck, 
a residual block (referred to as ``Res. Block") is applied, which constitutes the primary building block of the architecture. 

In the encoder, the left-hand side of Fig.~\ref{fig:architecture}, the input coefficients are divided into octave-specific ``pieces" and processed separately using ``In. Blocks". The features from each octave are then concatenated along the frequency dimension to the corresponding latent vectors of the U-Net at corresponding time resolutions, as represented graphically in Fig.~\ref{fig:architecture}. Additionally, residual connections are applied between the (resized) input features and the corresponding latent vectors of the U-Net to facilitate information flow through all layers of the encoder.

\begin{figure} [t]
    \centering
    \includegraphics[width=0.9\columnwidth]{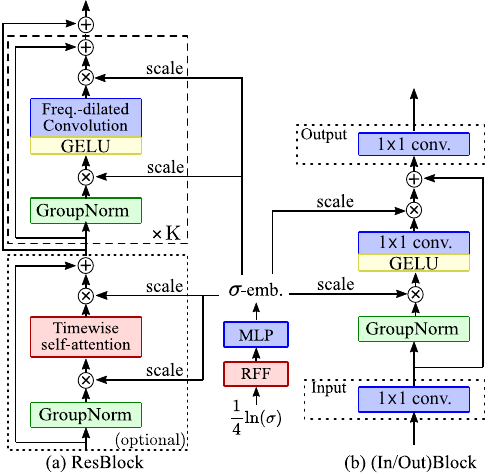}
    \caption{Building blocks of the backbone U-Net architecture, cf.~Fig.~\ref{fig:architecture}. }
    \vspace{-0.5cm}
    \label{fig:build_blocks}
\end{figure}

The decoder, or the right-hand side of Fig.~\ref{fig:architecture}, comprises the main signal path containing ``Res. Blocks" and the outer path with residual connections. As the temporal resolution is upsampled in the main path, the features from the lowest octaves at each layer are discarded and projected to the ``outer" path via ``Out. Blocks." In the outer path, at each resolution, the lowest octave is extracted and sent to the ICQT block, as indicated in the right part of Fig.~\ref{fig:architecture} lines.
This dual-path strategy is inspired by Karras et al.~\cite{karras2020analyzing}, and its purpose is to improve the gradient flow during the optimization process.

\subsubsection{Building Blocks of the Architecture}
The building blocks of Fig.~\ref{fig:architecture} are presented in detail in Fig.~\ref{fig:build_blocks}. They are all conditioned with the noise-level embedding $\sigma$-emb, which is built with random Fourier features (RFF) \cite{tancik2020fourier} followed by a multi-layer perceptron (MLP) having three layers. The conditioning is realized with feature-wise linear modulation \cite{perez2018film},  without shifts.
The ``In. Block" shown in Fig.~\ref{fig:build_blocks}(b) applies a 1$\times$1 convolution to expand the channel size from two (real and imaginary) to the required number of latent features at every layer. They are followed by Group Normalization (without shift operations), a Gaussian-error-linear-unit (``GELU") non-linearity, and a linear layer.
The ``Out. Blocks" have a similar form, but with the $1\times1$ convolution placed at the end, mapping the latent vector to a channel size of two.

Fig.~\ref{fig:build_blocks}(a) shows that each residual block, ``Res. Block", contains a stack of shift-free Group Normalization layers, followed by a ``GELU" non-linearity and convolutions in both time and frequency, but with exponentially-increasing dilations in the frequency dimension, meant to provide a wide receptive field while exploiting the symmetry of pitch-equivariance. 
We additionally include a timewise self-attention layer in the deeper ``Res. Block" layers, as explained in Sec.~\ref{selfatt}.


In contrast to our former work  \cite{moliner2022solving}, frequency-positional embeddings designed to encode absolute frequency positional information are not used. The reason is that, while the absolute frequencies cannot, in principle, be retrieved with a CNN, they can, in practice, be spuriously learned through the use of zero padding \cite{islam2020much}.
With this modified architecture, zero padding is used in the convolutional layers at each intermediate stage, also in the frequency dimension, propagating absolute positional information throughout the network, even at the shallower layers. 
We observed that, in this setting, the use of frequency-positional embeddings provided no significant benefit.



\subsubsection{Timewise Self-Attention} \label{selfatt}

The motivation behind using timewise self-attention is to allow the model to learn global features in the time dimension, overcoming the locality of CNNs. The use of attention would allow the model to analyze similarities between different segment pairs in time, a feature that could intuitively be highly beneficial for the task of inpainting. The latent features are two-dimensional (time and frequency), but since the idea is to apply attention only through time and not in frequency, some modifications are required with respect to the basic self-attention mechanism \cite{vaswani2017attention}.
\begin{figure}[t!]
    \centering
    \includegraphics[width=0.78\columnwidth]{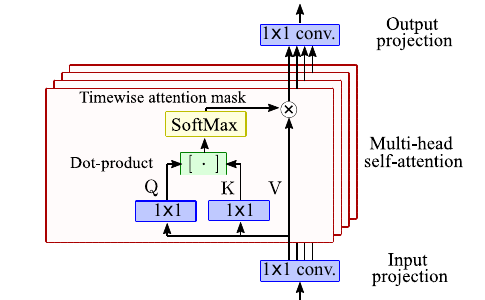}
    \caption{Timewise self-attention block used in  Fig.~\ref{fig:build_blocks}.} 
    \vspace{-0.5cm}
    \label{fig:attention}
\end{figure}

Fig.~\ref{fig:attention} presents the functionality of the timewise self-attention block.  
In order to reduce the otherwise unfeasible computational complexity, we introduce a $1\times1$ convolution before the self-attention mechanism. This reduces the number of channels, 
which can be quite large (up to 256), to only a few attention heads (set to eight in this work).
For each head, the queries $Q$ and keys $K$ are computed with a linear layer that sees the frequency dimension as the feature dimension. A timewise attention mask is computed via standard dot-product attention, and is later applied to the values $V$.
In order to preserve the structure in the frequency dimension, we do not apply any processing to the values $V$, apart from the timewise attention. Finally, another $1\times1$ convolution is applied at the output to expand the reduced number of heads to the original channel size. 

Note that, as implemented in this work, the use of timewise self-attention breaks the translation-equivariant property of fully convolutional networks, rendering the model unsuitable for processing sequences of different lengths than the one used during training. If one wishes to process a longer sequence, a segment-by-segment approach with a fixed segment length can be applied.

\subsubsection{Hyperparameter Specification}

The architecture of Fig.~\ref{fig:architecture} is designed to work at a sampling frequency of $f_\text{s}=44.1$\;kHz. We use a CQT with $B=64$ bins per octave and $N_\text{oct}=8$ octaves.
The depth of the U-Net matches the number of octaves, and the feature sizes range from 64 features at the shallower U-Net layers to 256 features at the bottleneck.
The number of stacked dilated convolutions on each ``Res.~Block" ranges from two to eight, with fewer dilations at the shallower layers since they need to cover fewer frequency bins and, thus, a large receptive field is not needed.
Because of their quadratic complexity, timewise self-attention is only used at the three deepest layers, where the time resolution has been significantly reduced. The total parameter count is 242 million parameters.
We refer to the public repository for further specifications\footnote{\href{https://github.com/eloimoliner/audio-inpainting-diffusion}{https://github.com/eloimoliner/audio-inpainting-diffusion/tree/main/conf}}.


\section{EVALUATION}
\label{evaluation}
The performance of the proposed method, which we refer to as CQT-Diff+, is evaluated for inpainting short to middle-sized gaps in musical recordings, ranging from 25\,ms to 300\,ms.
For comparison, two baselines are considered:
\begin{itemize}
    \item \textbf{LPC}: A method based on signal extrapolation using linear predictive coding \cite{kauppinen2002audio}.
    \item \textbf{A-SPAIN-L:} A sparsity-based method for audio inpainting with dictionary learning \cite{taubock2020dictionary}, which is regarded as a state-of-the-art method for short-gap inpainting.
\end{itemize}

\noindent Results of both objective and subjective experiments are reported. All experiments use the sample rate of 44.1\,kHz.
In the objective evaluation, we analyze various gap lengths within intervals of 25 ms. 

In the subjective evaluation, we examine four different gap sizes: 50\,ms, 100\,ms, 200\,ms, and 300\,ms. In all instances, we intentionally introduce four gaps uniformly across audio segments that have a duration of 4.17\,s. These gaps are applied simultaneously at predetermined time locations.
Other machine-learning-based methods could not be included in the evaluation, since they were either not designed for wideband music audio \cite{ebner2020audio, pascual2021adversarial, moliner2022solving}, or they do not allow enough flexibility to be tested with gaps of different length \cite{marafioti2019context, marafioti2020gacela}.
For instance, our previous diffusion model  \cite{moliner2022solving}, which corresponds to a prior iteration of the proposed method, could not be included as a baseline either, being unsuitable to work at sample rates higher than 22.05\,kHz due to memory constraints. 

The initial hypothesis is that our method provides no advantage against the baselines when the gap is very short, as, in this case, stationary conditions can be safely assumed. However, as the duration of the gap increases, the problem gets more challenging, and the performance of the baselines is likely to degrade. On the other hand, a diffusion-based generative model should not suffer from this limitation and should be capable of generating audio content regardless of the gap length.
Thus, the question we want to resolve is the following: \emph{
How does the performance of CQT-Diff+ compare to the baselines in terms of reconstruction quality as the gap length increases?}

\subsection{Training}

We train our model with the MusicNet dataset, a collection of 330 freely-licensed classical music recordings sampled at 44.1\,kHz. 
MusicNet is a multi-instrument dataset containing recordings from a wide variety of acoustical environments and recording conditions, representing a challenging and realistic scenario.
We use a segment length of 4.17\,s, limited by memory requirements.
The training is performed using the Adam optimizer,
with a learning rate of $2\times10^{-4}$ and a batch size of four. The model is trained for roughly 500,000 iterations, taking approximately five days on a single NVIDIA A100 GPU.
During training, we track an exponential moving average of the weights, which corresponds to the one used during testing.


\subsection{Objective Evaluation} 




We first conduct an objective evaluation where we report three metrics.
The first one is \emph{log-spectral distance} (LSD) \cite{gray1976distance}, a reference-based metric specified as
\begin{equation}
   \text{LSD}=\frac{1}{T} \sum_{t=1}^T \sqrt{
   \frac{1}{K}
   \sum_{k=1}^K
   \left(
   \log |X_{t,k}|^2- \log |\hat{X}_{t,k}|^2
   \right)^2
   },
\end{equation}
where $X_{t,k}=\text{STFT} (\mathbf{x}_0)$ and $\hat{X}_{t,k}=\text{STFT} (\mathbf{\hat{x}}_0)$  are the STFTs of the reference $\mathbf{x}_0$ and the restored audio signal $\mathbf{\hat{x}}_0$, respectively. For the STFT computation, an analysis window of $K=2048$ samples and a hop length of 512 samples is used. 
The LSD provides information about the reconstruction performance, with respect to the original signal. 

We also use the \emph{Objective Difference Grades} (ODG), estimated using the PEMO-Q auditory model \cite{huber2006pemo}. This metric is also reference-based and aims to replicate the subjective difference grades that are obtained through a subjective listening test.
The last metric is the reference-free \emph{Fréchet Audio Distance} (FAD) that compares the statistics of a set of generated data against those of a reference dataset \cite{kilgour2019frechet}. This metric has been demonstrated to correlate with perceptual audio quality \cite{kilgour2019frechet}. In our case, we compare the distribution of inpainted audio signals with that of the original ones.

\begin{figure*}
    \centering
    \includegraphics[width=0.75\textwidth, trim={0 0.25cm 0 0.25cm}, clip]{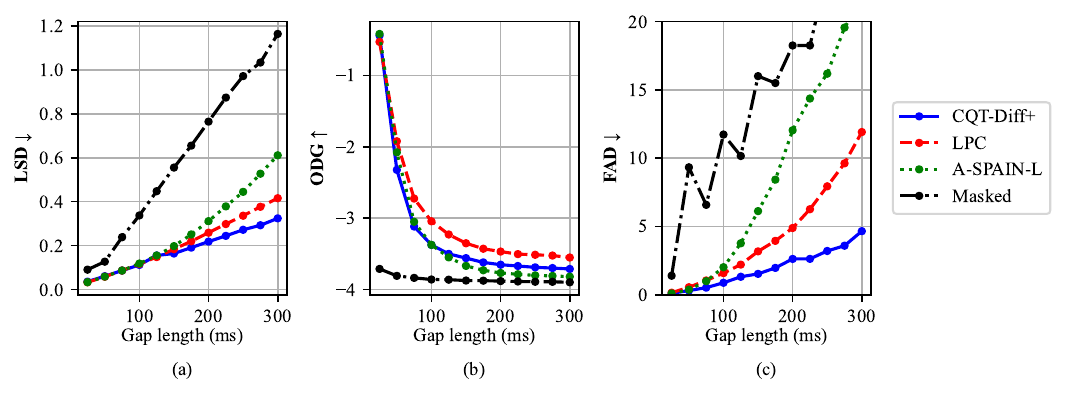}
    \caption{Average objective metrics, including (a) log-spectral distance (LSD), (b) Objective Difference Grades (ODG), and (c) the Fréchet Audio Distance (FAD), computed for various gap lengths from 25 to 300\,ms.
    Lower is better for LSD and FAD, whereas higher is better for ODG. 
    The proposed method (CQT-Diff+) obtained competitive results against the baselines in the reference-based metrics LSD and ODG, while being superior in terms of LSD.
    }
    \label{fig:objectivemetrics}
\end{figure*}

The results for different gap lengths are plotted in Fig.~\ref{fig:objectivemetrics}. In our evaluation, we used a subset of the MusicNet test set \cite{thickstun2016learning} comprising 60 randomly selected 4.17-s samples, each of them containing four equally-spaced gaps. The test samples were not seen during the training of the proposed method.
We deliberately chose a smaller test set due to computational limitations. We believe that expanding it would not notably alter the results, considering the added computational load.


Fig.~\ref{fig:objectivemetrics}(a) shows that, according to the LSD metric, for gaps smaller or equal to 100\,ms, the proposed method yields a performance similar to the baselines and marginally outperforms them for longer gaps.
The results of the ODG metric are presented in Fig.~\ref{fig:objectivemetrics}(b), showing how all methods obtain similar values for small gap lengths, with LPC performing marginally better.
Above 100\,ms, all the ODG values are below $-3$, which refers to annoying or very annoying sound quality \cite{ITU1387}.
Finally, Fig.~\ref{fig:objectivemetrics}(c) shows the FAD results, where lower values ($<$ 5) indicate that the distribution of inpainted audio is statistically similar to the reference.
The proposed method consistently achieves lower FAD values than the compared baselines,
meaning that the inpainted audio is in-distribution with the rest. On the other hand, the baselines show a strong decline in terms of FAD as the gap size increases. 

\begin{figure*} [t]
    \centering
    \includegraphics[trim={0 0.25cm 0 0.2cm},clip, width=0.78\textwidth]{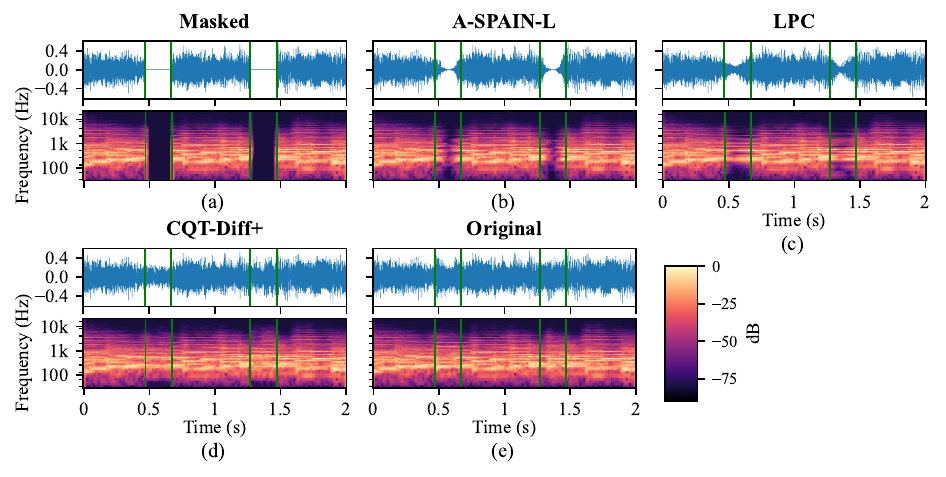}
    \vspace{-0.2cm}
    \caption{Audio inpainting examples with two gaps of 200\,ms, for all five signal types included in the listening test, including waveform and spectrogram representations. The start and end timestamps of the gaps are highlighted with green lines. The masked (a) and the original waveforms (e) are also included for comparison.
    The proposed method, CQT-Diff+ (d), produces more coherent and realistic reconstructions than the compared baselines (b), (c).
    }
    \label{fig:examples}
\end{figure*}

 \subsection{Subjective Evaluation}
\begin{figure*}
    \centering
    \includegraphics[width=0.78\textwidth]{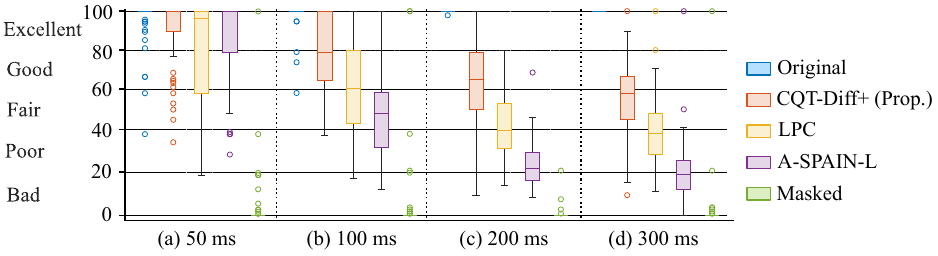}
    \vspace{-0.2cm}
    \caption{Boxplot diagrams representing the results of the subjective listening experiment for evaluated gap lengths of 50 ms (a), 100 ms (b), 200 ms (c), and 300 ms (d).}
    \label{fig:subjectivemetrics}
\end{figure*}

Since there is no guarantee that objective metrics provide reliable information about the perceived quality of the inpainting methods, we conducted a subjective listening experiment.
The listening test was designed in accordance with the MUSHRA recommendation \cite{ITURmushra}, using the webMUSHRA evaluation tool \cite{schoeffler2018webmushra}.
The participants were asked to rate, on a scale from 0 to 100, the perceptual similarity of each item with respect to the reference, which was the original audio sample (without gaps).
 
The conditions included a low anchor (a masked version of the reference with four gaps); three reconstructed versions of the low-anchor, produced using the LPC, A-SPAIN-L, and the proposed CQT-Diff+; and a hidden reference (the original unprocessed signal). Fig.~\ref{fig:examples} shows an example of all five conditions with two gaps. The listeners were allowed to loop and focus in detail on the gap locations.  The items represented four gap lengths (50, 100, 200, and 300\,ms) and 12 randomly-picked 4.17-s examples from the MusicNet test set.
The test contained a total of 48 pages of the five items above to be evaluated.
In order to reduce the duration of the experiment and avoid listening fatigue, the test was split into two equal-length parts of 24 pages that were alternatively assigned to the participants.
A total of 15 volunteers, all without reported hearing defects, participated in the experiment. The average age of the listeners was 28 years. The audio examples used for the listening test are available at the companion webpage\footnote{\href{http://research.spa.aalto.fi/publications/papers/jaes-diffusion-inpainting/}{http://research.spa.aalto.fi/publications/papers/jaes-diffusion-inpainting/}}.


The results of the listening test are presented in Fig.~\ref{fig:subjectivemetrics}. 
Except for the 50\,ms case where A-SPAIN-L was superior, LPC obtained higher scores than A-SPAIN-L.
For the gap length of 50\,ms, the proposed method obtained scores similar to the compared baselines, all of them close to 100.
For the remaining evaluated gaps longer than 50\,ms, the proposed method outperformed the baselines.
We studied the statistical significance of the score differences between CQT-Diff+ and LPC through a Wilcoxon signed-rank test, that gave a p-value of 1.2$\times 10^{-4}$, 1$\times 10^{-9}$, and 3.5$\times 10^{-9}$
for the gap lengths 100\,ms, 200\,ms, and 300\,ms, respectively. We conclude that the differences are significant since the p-values are way below 0.05. 
With the exception of gap lengths of 50\,ms, the findings depicted in Fig.~\ref{fig:subjectivemetrics} reveal that the proposed approach consistently achieves median scores ranging from 50 to 80. These scores exhibit a proportional decrease as the gap length increases. In the case of the shortest 50\,ms gaps, the median score for the CQT-Diff+ method reaches 100, indicating that it was difficult for listeners to find discerning differences in this particular test scenario.

Considering the test question, the listening test result can be interpreted so that the proposed diffusion model performs at least as well as the compared baselines for all gap lengths.
The minimum gap length for which the reconstruction using the proposed method is better than the baselines is 100\,ms, above that CQT-Diff+ consistently outperforms the baselines.
We can conclude that, up to the length of 200\,ms, the proposed CQT-Diff+ algorithm produces perceptually ``good'' audio inpainting (median scores above 60), although distinguishable from the reference in pairwise comparison. For the gap length of 300\,ms, the proposed method offers ``fair'' sound quality.


To gain a deeper understanding of the subjective test results, we qualitatively analyze a specific example depicted in Fig.~\ref{fig:examples}. This figure exhibits waveform and spectrogram representations of a masked music signal in Fig.~\ref{fig:examples}(a), along with three reconstructed versions, Figs.~\ref{fig:examples}(b), (c), and (d), and the original signal for reference in Fig.~\ref{fig:examples}(e), all with a gap length of 200\,ms.
Upon examination, A-SPAIN-L evidently generates an attenuated reconstruction that fades out towards the middle of the gap and fades in again before reaching the gap's end.
In practice, the method shortens the dropout but cannot fill it.
 The decay observed in the reconstruction arises due to the sparsity penalty that restricts the generation of content further into the gap.
On the other hand, relying on extrapolation, LPC suffers less from this issue. Nevertheless, this method is only capable of extending stationary sounds and cannot create new attacks and events. Consequently, the reconstructions produced by LPC often sound artificial.

Visually, the reconstruction generated with the proposed CQT-Diff+ algorithm fills the gap in a credible manner in Fig.~\ref{fig:examples}(d). However, the comparison with the original signal shown in Fig.~\ref{fig:examples}(e) reveals that the reconstruction of the proposed methods is not exact, since the two waveforms look different.

\section{CONCLUSION} 
\label{conclusion}

This paper presents a novel audio inpainting method CQT-Diff+ that is based on recent diffusion models. For the reconstruction of short gaps of 50\,ms or less, the proposed method works as well as a previous high-quality inpainting method. For longer gaps, from 100\,ms to 300\,ms, the CQT-Diff+ method outperforms the baseline algorithms and retains good or fair quality. 

One limitation of the method presented in this paper is that the performance is limited to the types of audio recordings seen during the training, in this case, classical music.
To demonstrate the versatility of the proposed approach, we also report, in the form of audio examples in the companion webpage, results obtained with a model trained on a large variety of sound effects. However, these results were excluded from the evaluation in this paper.
In the future, the generalizability of the diffusion-based audio inpainting technique should be evaluated by considering models trained on a wider variety of audio recordings.

Another promising direction for future work could involve providing the CQT-Diff+ model with conditional information.
In the results of our tests reported elsewhere, a generative method based on a diffusion model created plausible content and new events, when the gaps were 1.5\,s long \cite{moliner2022solving}.
However, due to the lack of contextual information, the results were difficult to control. 
A conditional diffusion approach may offer a higher degree of control over the results, especially when dealing with extremely long gaps.


    



\section{ACKNOWLEDGMENT}
This research is part of the activities of the Nordic Sound and Music Computing Network (NordForsk project no.~86892). We acknowledge the computational resources provided by the Aalto Science-IT project. We thank the volunteers who participated in the listening test. We are grateful to Luis Costa for proofreading.

\bibliography{jaes.bib}
\bibliographystyle{jaes.bst}

\appendix


 \biography{Eloi Moliner}{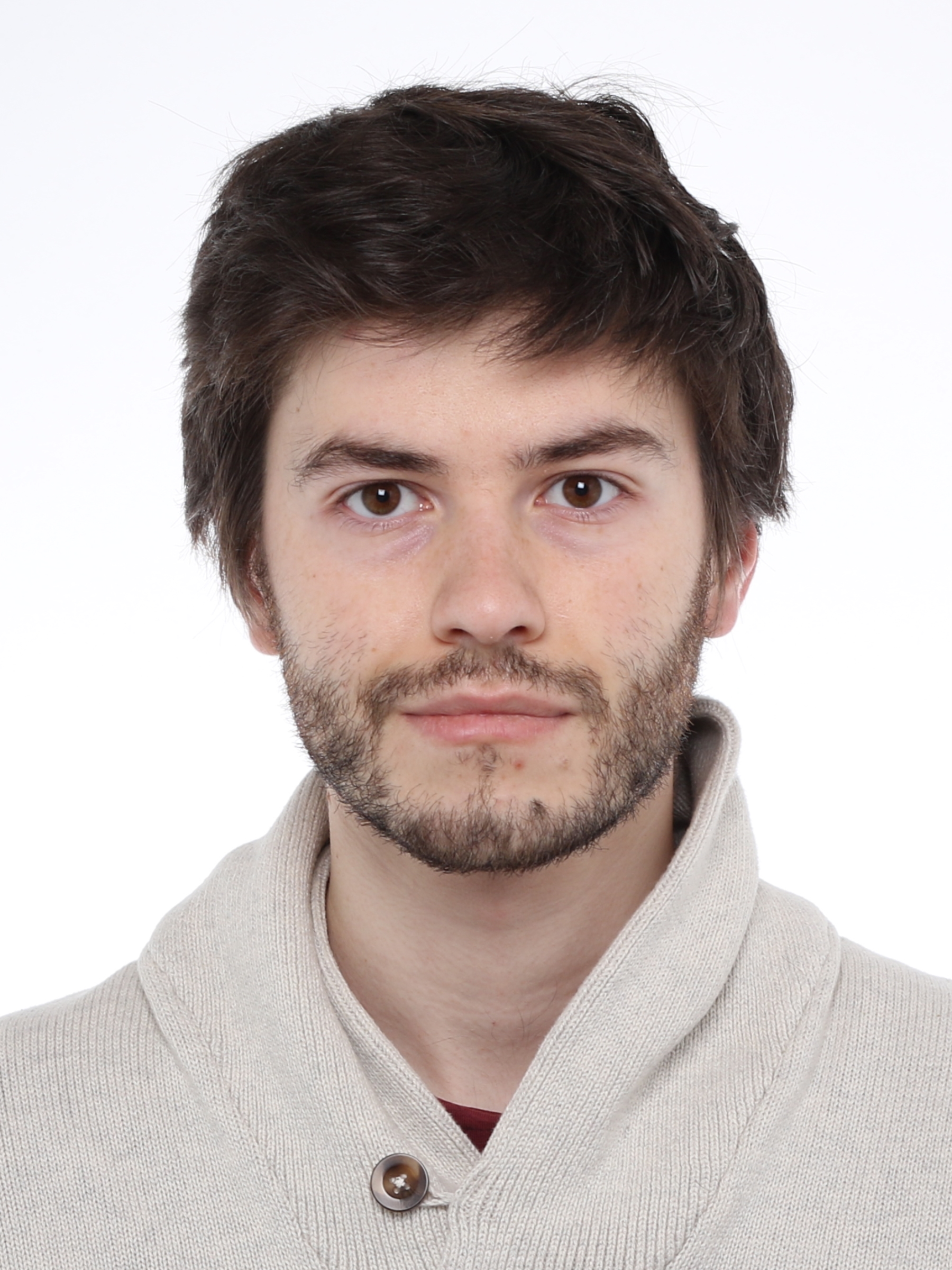}{Eloi Moliner received his B.Sc. degree in Telecommunications Technologies and Services Engineering from the Polytechnic University of Catalonia, Spain, in 2018 and his M.Sc. degree in Telecommunications Engineering from the same university in 2021.
He is currently a doctoral candidate at the Acoustics Lab of Aalto University in Espoo, Finland. His research interests include digital audio restoration and audio applications of machine learning. He is the winner of the Best Student Paper Award of the 2023 IEEE ICASSP conference.}
 
\biography{Vesa Välimäki}{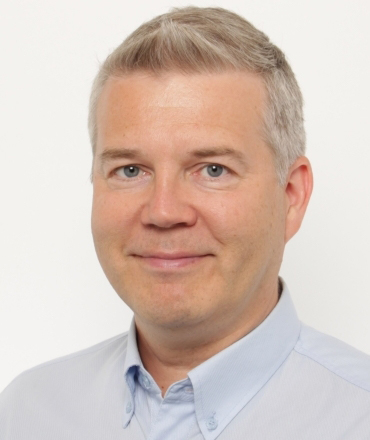}{Vesa V\"alim\"aki is Full Professor of audio signal processing and Vice Dean for Research at Aalto University, Espoo, Finland. He received his D.Sc. degree from the Helsinki University of Technology in 1995. In 1996, he was a Postdoctoral Research Fellow at the University of Westminster, London, UK. In 2008--2009, he was a visiting scholar at  Stanford University. 
He is a Fellow of the AES, the IEEE, and the Asia-Pacific Artificial Intelligence Association. Prof.~V\"alim\"aki is the Editor-in-Chief of the Journal of the Audio Engineering Society.}
\end{document}